\documentclass[preprint,authoryear,12pt]{elsarticle}

\def \eq {\begin{equation}}
\def \qe {\end{equation}}
\def \eqa {\begin{eqnarray}}
\def \aqe {\end{eqnarray}}
\def \spc {\hspace{2pt}}
\DeclareMathSizes{12}{12}{10}{6}  
\newcommand{\comment}[1]{}
\usepackage{amssymb}
\begin{document}
\begin{frontmatter}
\title{Visual illusion due to the interaction of flickering and acoustic vibrotactile signals}
\author{Ruggero Micheletto\corref{rug}}
\ead{ruggero@yokohama-cu.ac.jp}
\author{Maria Fernanda Avila Ortega\fnref{fnFer}}
\ead{maria-o\_o@i.softbank.jp}
\address{Yokohama City University, Graduate School of Nanobioscience, Department of Nanosystem Science, 22-2 Seto Kanazawa-ku, 236-0027 Yokohama, Japan}
\cortext[rug]{Telephone +81.45.787.2167}
\fntext[fnFer]{First names: "Maria Fernanda", telephone +81.80.4090.5457}
\begin{abstract}
We studied the influence of mechanical vibrotactile signals in the acoustic range to the visual perception of flickering images. These  images are shown on a CRT screen intermittent at about 75 Hz, without external perturbations are perceived as constant and stable. However, if presented together with a controlled acoustical vibration an illusion is perceived. The images appears to float out of the screen, while the rest of the room is still perceived normally. The acoustical signal given to the subjects were of very low frequency (below 100Hz) and low amplitude (almost inaudible). The stimuli were transmitted through direct contact to the subject's chin with the use of a plastic stick connected to a speaker. The nature of the illusion is described and a basic theoretical model is given.
\end{abstract}
\end{frontmatter}
\newpage
Vibrations are everywhere. When we walk, all our body shakes. Our eyes are not exception, they jitter while we move. As a consequence, images projected on our retina are trembling, moving chaotically in random directions (\cite{mura:1998}). 

Moreover, we are continuously scanning what we are looking at, in jerky criss-cross movements that are called “saccades”. So, even in the absence of external mechanical perturbations, the images on the retina are randomly shifting because of the saccades. Despite of this, the brain is able to remove spurious movements and produce the perception of clean and stable images in any condition (\cite{verc:1984}).

This "cleaning up" elaboration process is a heavy and complex task, however we are not aware of what is happening, nor we feel any effort. It is a process that is unconscious and transparent to us (\cite{vela:1997,peli:2003}).

In this study we perturb this process to make it surface to the conscious level in order to investigate its characteristics. We describe a simple experiment that, interfering with the the visual elaboration process, induce errors that can be perceived at the cognitive level. The method combine the interaction of an acoustic vibrotactile signal and the perception of flickering images. 

To generate the flickering, we used a cathodic ray tube monitor (CRT). This once common device draws images point by point with an electronic brush that scan the screen at a frequency of 75Hz. 

\begin{figure}[h]
\begin{center}\leavevmode
\includegraphics[width=6.5cm]{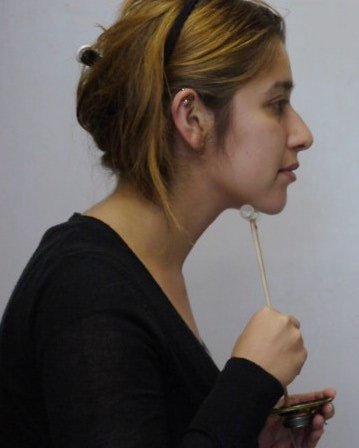}
\caption{The very simple way we applied the acoustical perturbation to the subject retina. The vibration is in the 50$\sim$100 Hz range and of very low intensity (almost inaudible)}
\label{fg:ferdinanda}\end{center}\end{figure}

We setup the following apparatus: a controlled acoustical vibration of about same frequency as the display is transferred to the subject eye by a small wooden stick glued to a round plastic cylinder that is placed horizontally under the chin as shown in figure \ref{fg:ferdinanda}. The stick is mechanically connected to a speaker that the subject simply holds in his hand. Because of its low amplitude and frequency the vibration is barely audible and does not disturb the subject, however the sound is efficiently transmitted to the whole head. 

\begin{figure}[h]
\begin{center}\leavevmode
\includegraphics[width=8.5cm]{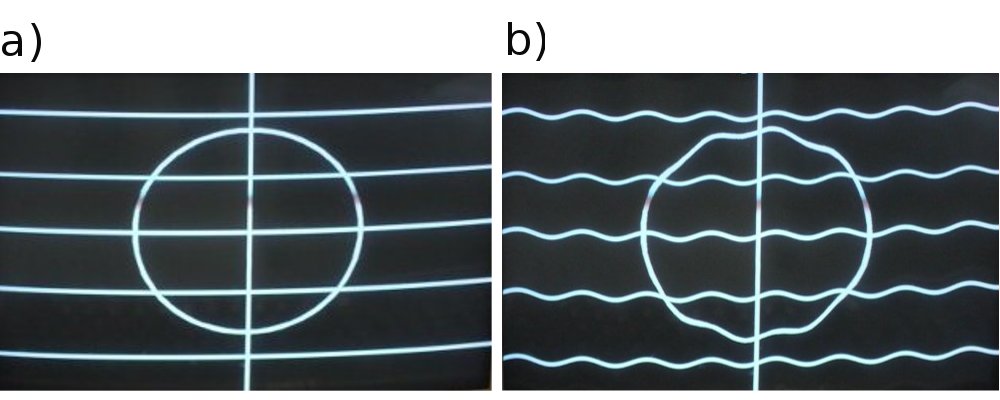}
\caption{Panel a). The image presented to the subjects. Panel b). A representation of the perceived illusion: a wavy distortion of the picture that appears also to dynamically float off screen}
\label{fg:screens}\end{center}\end{figure}

The frequency is generated by a sinusoidal function generator while the subject is seated comfortably looking at a geometric image on the screen (see the image in figure \ref{fg:screens}a). 

While varying the frequency and amplitude of the sound, we asked the subjects to say when the perception was becoming unstable.

When the audio signal reaches a frequency close to that of the screen flickering, something interesting happens. While the rest of the room appears unchanged to the subject, the images presented on the CRT screen begin to move and appear unstable. The images perceived are distorted with perturbations that change in strength and shape in line with the audio signal intensity and frequency (see figure \ref{fg:screens}b). 
\begin{figure}[h]
\begin{center}\leavevmode
\includegraphics[width=8.5cm]{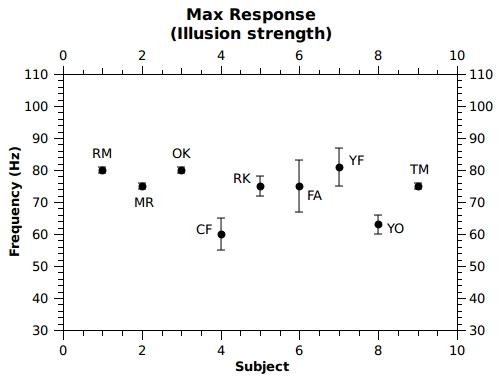}
\caption{The frequencies that resulted in the maximum illusion strength for several subject. Some subjects performed the experiments twice or three times, for them the error bar is given}
\label{fg:freqs}\end{center}\end{figure}

\begin{figure}[h]
\begin{center}\leavevmode
\includegraphics[width=8.5cm]{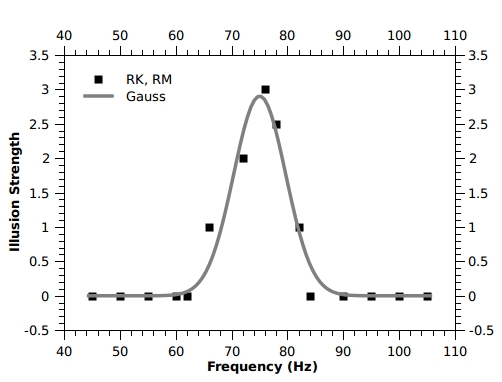}
\caption{The strength of the illusion is measured by a subject in terms of four subjective levels: maximum, medium, weak and no illusion. To plot the data a value is assigned to each of the level as following: maximum=3, medium=2, weak=1 and zero for no illusion. The effect appears when the acoustical perturbation is comparable with the flickering frequency and decrease when it exceed it with a Gaussian profile as shown. Blocks are the average values for subject RM and OK and continuous curve is the fit.}
\label{fg:illstr}\end{center}\end{figure}

We asked the subjects to report 4 levels of intensity to quantify this subjective phenomenon: no illusion (image perfectly stable), weak, medium and maximum. In figure (\ref{fg:freqs}) are plotted the frequencies of maximum effect for different subjects. The illusion appears and decays with a Gaussian profile as shown in the plot (\ref{fg:illstr}). 

It is worth repeating that all other objects in the room are perceived as perfectly still, only the images on the display appears to oscillate (a schematic representation of the percept is in figure \ref{fg:screens}b).

What is the cause of the illusion? If the retinal persistence is the only reason why we perceive stable images from the flickering stimuli, even the vibration introduced by the external acoustical disturbance, that have similar frequency, would average out and have no effect, but this is not the case.

In other words we cannot assign the illusion to the sole effect of the mechanical vibration that perturbs the retina, because all the other objects surrounding the display are perceived normally. It the interaction of the two causes (flickering and vibrotactile signal) that give rise to the illusion. \par To have an insight about this, we use a simplified treatment. We define the image on the retina as a two-dimensional map $$ M=\rho^{\spc xy}(t) $$ and the perceived pattern: $$ \tilde{M}=\tilde{\rho}^{\spc xy}(t),$$ where $\rho$ represents the intensity of the point and the $xy$ apex indicates that the parameter is a bidimensional map in the $xy$ coordinates (from now-on the tilde always indicates perceptions). We then represent the external acoustic signal disturbance as perturbation of pulsation $\omega_o$, in this formalism the general expression relating patterns to percept is :

 \eq \tilde{M} =  T^{w_o}[M]. \nonumber\\ \qe

The function $T$ is the unknown transfer function that represents the complex algorithm that the brain uses to generate the perception from pattern $M$. The apex $w_o$ represents its dependence on the external vibrotactile signal frequency.

Clearly, we do not know the exact form of $T$, however we can derive relevant characteristics of it from the illusion. We consider two optical stimuli, the fix objects external to the CRT screen $$ M_e=\rho_e^{\spc xy} $$ and the flickering images presented by the screen (for simplicity we assume all screen pixels blink simultaneously)$$ {M_s}=\rho_s^{\spc xy}\cos (\omega_s t)$$ the indexes $e$ and $s$ represent the external and screen stimuli respectively.
When we perturb the visual system with an acoustic frequency $\omega_o$, if the stimulus on the screen have a different pulsation $\omega_s$, the visual system has the ability to average everything out. So $T$ have this form:
$$ \tilde{M}=T^{\spc \omega_s\neq\omega_o}\left[ \rho^{\spc xy}_s\cos (\omega_s t)\right] =\rho^{\spc xy}_s, $$
 the percept is a constant pattern as seen experimentally before 60Hz or after 85Hz in figure \ref{fg:illstr}.
However, when the two pulsations are comparable, the illusion prevail and subjects perceive fluctuating images (peak in figure \ref{fg:illstr}). The percept is not stable, oscillating with pulsation $\Omega$: $$ \tilde{M}=\rho^{\spc xy}_s \cos(\Omega t),$$ we can write: 

$$ \tilde{M}=T^{\spc \omega_s\approx\omega_o}\left[ \rho^{\spc xy}_s \cos (\omega_s t) \right]= \rho^{\spc xy}_s cos(\Omega t) $$

naturally $\Omega$ depends on the frequency of the external vibrotactile perturbation. It is a function of $w$ different from zero only when the acoustical perturbation and the flickering stimulus frequencies are similar. The shape of $\Omega$ is in fact deducible from experiments and coincide with the profile of the curve of fig. \ref{fg:illstr}. This can be modeled with a Gaussian function as:

$$
\Omega=(\omega-\omega_o)e^{-(\omega-\omega_o)^2/\delta^2}  ,
$$

where $\delta$ is a parameter proportional to the spread of the curve. This value can be estimated from out data as $ \delta=10Hz$. 

\begin{figure}[h]
\begin{center}\leavevmode
\includegraphics[width=8.5cm]{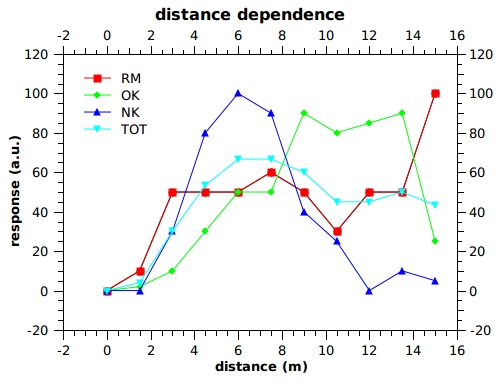}
\caption{The strength of the illusion in function of distance. The screen was placed at increasing distances and the subjects were asked to give a relative evaluation of the illusion strength in percentage. TOT symbol represent the average of all subjects.}
\label{fg:distance}\end{center}\end{figure}

We studied systematically the distance dependence and found that this visual effect tends to be more prominent ad distance and to decay in close proximity of the screen. Data in figure \ref{fg:distance} show that at very close distances the illusion is not perceived. At these ranges, image covers a wider area on the retina, the number of receptors involved is high enough for the brain to process the image avoiding any confusion. On the other hand, the illusion gets more dominant at distance for the opposite reason: the image on the retina gets smaller, involving less cells; the brain has less information to process and the flickering disturbance has an easier task to prevail. We would expect a monotone increase in the illusion strength at greater distances, however as seen in figure \ref{fg:distance}, some subject exhibit a decay due to intervening visual resolution problems. 
\par
\begin{figure}[h]
\begin{center}\leavevmode
\includegraphics[width=8.5cm]{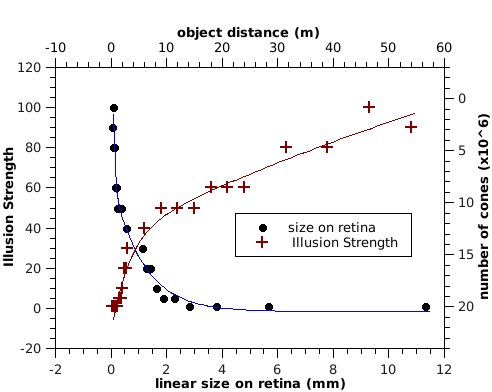}
\caption{The strength of the illusion in function of the number of receptors involved. As the number of receptors drops, the Brain loose its ability to stabilize the flickering image giving rise to the illusion.}
\label{fg:strength}\end{center}\end{figure}

To have a confirmation of the phenomenon, we modelled the the eye as a perfect lens, with the retina placed at focal distance length L=17mm. Considering that the pattern stimulus is about P=20cm in size, simple triangulations allow us to calculate the size of the image on the retina as $h_r=L*\frac{P}{d}$ where $d$ is the distance of the stimulus from the subject eye. Supposing that the perception occurs within the fovea and taking in account that typical cone density is of about 150.000 cones/mm$^2$, we plot the number of receptors involved against the distance. These data compared with the illusion strength shows that when the number of receptors drops, the illusion start to kick in suggesting that the brain algorithm robustness depend on how many receptors are involved in the elaboration, see figure \ref{fg:strength}.    

If the reader is curious to test this optical illusion by himself, he does not need a complicated setup. He just needs to find an old cathodic TV or monitor, switch it on and place himself at about 2-3 meters from it. To perceive the floating display illusion, it is not necessary a calibrated perturbation of the right amplitude and frequency as done in this study. The easiest way to perceive the illusion is simply eating a cracker or a biscuit few meter away from the blinking image. Crunching the biscuit will stimulate a broad spectrum of frequencies, also those in the correct range, giving rise to the effect. You will notice that the images on the TV appear to float out of the screen, while all the rest of the room appears unaffected. We are sure that this optical illusion has been noticed by most people with a sufficient level of attention in casual moments of everyday life. 
\par
The experiments presented show that the algorithm used by the visual system to stabilize images from saccades and external vibrations, can be disrupted and fail in presence of blinking images. Using a controlled vibrotactile mechanical stimulus on the subject, we found that this happens at frequency ranges similar to the one of the image, even if it is not perceived as blinking in normal conditions. Our study shows also that the phenomenon is dominant at longer distances when solid angle is smaller and less visual receptors are involved. We give a simple theoretical model that finds the explicit Gaussian relation between image blinking frequency and the vibrotactile stimuli. We believe that further study of this effect may lead to a better understanding of the relation between retina vibration stimuli and visual perception, shed light on the mechanism related to the dynamic formation of images and maybe help the development of methodology to discover visual perception related diseases.

\end{document}